\documentclass[aps,prl,longbibliography,twocolumn,superscriptaddress,amsfont,graphicx,nofootinbib,preprintnumbers]{revtex4-1}%
\UseRawInputEncoding
\usepackage{color,graphicx,epsfig}
\usepackage{ifpdf}
\usepackage{amsmath}
\usepackage{bm}
\usepackage[english]{babel}
\usepackage{amssymb}
\usepackage{braket}
\usepackage{setspace}
\allowdisplaybreaks[4]

\usepackage{hyperref}
\usepackage{enumerate}
\usepackage{lipsum}

\usepackage{slashed}

\usepackage{changes}

\begin{document}

\title{Plasmon-enhanced Direct Detection Method for Boosted sub-MeV Dark Matter}%


\author{Zheng-Liang Liang}
\email{liangzl@mail.buct.edu.cn}
\affiliation{College of Mathematics and Physics, Beijing University of Chemical Technology, Beijing, 100029, China}

\author{Liangliang Su}
\email{liangliangsu@njnu.edu.cn}
\affiliation{Department of Physics and Institute of Theoretical Physics, Nanjing Normal University, Nanjing, 210023, China}

\author{Lei Wu}
\email{leiwu@njnu.edu.cn}
\affiliation{Department of Physics and Institute of Theoretical Physics, Nanjing Normal University, Nanjing, 210023, China}

\author{Bin Zhu}
\email{zhubin@mail.nankai.edu.cn}
\affiliation{School of Physics, Yantai University, Yantai 264005, China}

\date{\today}

\begin{abstract}
Plasmon, a collective mode of electronic excitation in solid-state detectors, provides a novel way to detect light dark matter (DM). In this work, we present the conditions of DM to produce plasmon resonance, requiring relativistic velocities for light DM, and generalize the collective excitation framework to account for relativistic DM. As a demonstration, we consider the cosmic ray boosted DM (CRDM) and find that the plasmon resonance can be significantly enhanced in the scenario with a light mediator. Utilizing the first data from SENSEI experiment with the skipper-CCDs at SNOLAB, we obtain a new strong limit on the sub-MeV DM-electron scattering cross section.  

\end{abstract}

\maketitle

{\it Introduction.} Dark matter, constituting approximately 85\% of the Universe's mass, plays a crucial role in the structure formation and evolution of our Universe. Despite the overwhelming astrophysical and cosmological evidence confirming its existence, the nature of dark matter remains elusive. The Weakly Interacting Massive Particle (WIMP) hypothesis has been a leading candidate, yet recent null results from WIMP searches~\cite{PandaX-II:2020oim, LZ:2022ufs, XENON:2023sxq} have prompted extensive exploration of alternative possibilities, particularly focusing on sub-GeV dark matter~\cite{Essig:2011nj, CDEX:2019hzn, Emken:2019tni, Essig:2019xkx, Trickle:2020oki, Andersson:2020uwc, Vahsen:2021gnb, Su:2021jvk, Catena:2021qsr, Elor:2021swj, Chen:2022pyd, Catena:2022fnk, Li:2022acp, Wang:2023xgm,Bhattiprolu:2023akk}. Given that the nuclear recoil of light dark matter falls below the threshold of conventional detectors, great efforts have been devoted to utilizing ionization signals to extend sensitivity in direct detection experiments.

The sub-GeV DM with the coupling to the electrons can be detected in semi-conductor targets, such as SENSEI~\cite{SENSEI:2020dpa}, DAMIC~\citep{DAMIC:2019dcn}, EDELWEISS~\cite{EDELWEISS:2020fxc}, CDEX~\cite{CDEX:2022kcd} and SuperCDMS~\cite{SuperCDMS:2020ymb}. The density functional theory (DFT) method was first introduced in the description of the DM-electron in semi-conductors~\cite{Essig:2015cda}, and have been further extended to a broader range of target materials~\cite{Hochberg:2015pha,Essig:2016crl,Hochberg:2016ntt,Derenzo:2016fse,Knapen:2017ekk,Griffin:2018bjn,Trickle:2019ovy,Kurinsky:2019pgb,Trickle:2019nya,Campbell-Deem:2019hdx,Coskuner:2019odd,Geilhufe:2019ndy,Griffin:2020lgd,Prabhu:2022dtm,Esposito:2022bnu}. Other methodologies are still under development~\cite{Liang:2018bdb,Griffin:2021znd,Mitridate:2021ctr,Kahn:2021ttr,Trickle:2022fwt,Dreyer:2023ovn}. For example, the dielectric function that accounts for the screening has been incorporated into the formalism describing the DM-electron interaction in semi-conductor targets~\cite{Hochberg:2021pkt,Knapen:2021run,Knapen:2021bwg,Liang:2021zkg}. On the other hand, a collective excitation mode of the valence electrons in solid materials known as the plasmon has been considered as a possible explanation of the excess observed in several semi-conductor detectors~\cite{Kurinsky:2020dpb,Kozaczuk:2020uzb} and included in the direct detection ~\cite{Gelmini:2020xir,Knapen:2021run}.

\begin{figure}[ht]
  \centering  
\includegraphics[height=6.5cm,width=8.7cm]{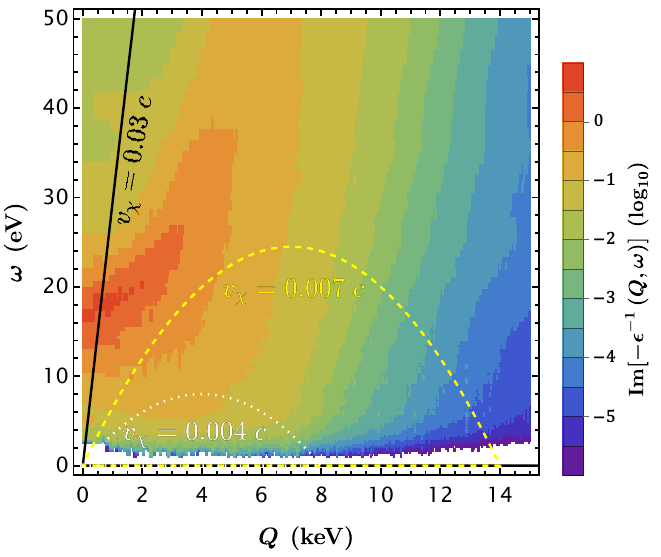}
\caption{The energy loss function $\mathrm{Im}\left[{-\epsilon^{-1}\left(\mathbf{Q},\omega\right)}\right]$  for silicon semi-conductors on the plane of $Q$ and $\omega$, which is calculated with the DFT methods~\cite{Knapen:2021run,Knapen:2021bwg}. We also depict the kinematically accessible phase spaces for a $1~\mathrm{MeV}$ DM with the incident velocities $v_{\chi}= 0.004c$, $0.007c$, and $0.03c$, each bounded by white dotted, yellow dashed and black solid lines, respectively. In 
our work, we adopt the natural units $c = \hbar=1$.}
\label{fig:ELF}
\end{figure} 
 
In this work, we investigate the plasmon excitation induced by the interaction between the accelerated DM and the semi-conductor detectors. One key observation is that if light DM moves fast enough, it can excite the plasmon in silicon and germanium detectors. The resonance behavior of the plasmon is described by the energy loss function, which is defined as the imaginary part of the inverse dielectric function, $\mathrm{Im}\left[{-\epsilon^{-1}\left(\mathbf{Q},\omega\right)}\right]$ and measures the response of the target material to an external perturbation.

To illustrate this, in Fig.~\ref{fig:ELF}, we show the energy loss function for silicon on the plane of the three-momentum $Q$ ($Q\equiv|\mathbf{Q}|$) and the energy $\omega$ transferred to the target. A resonance structure corresponding to the plasmon excitation is seen in the region of $Q<3~\mathrm{keV}$ and $\omega\sim15~\mathrm{eV}$. Besides, we also present the kinematic relationship $\ensuremath{E_{\chi}=\sqrt{\left(p_{\chi}-Q\right)^{2}+m_{\chi}^{2}}+\omega}$ for a $1~\mathrm{MeV}$ DM with the incident velocities $v_{\chi}= 0.004c$, $0.007c$, and $0.03c$. We observe that the minimum incident velocity $v_{\chi} \sim 10^{-2} c$ is required to excite the plasmon in the silicon targets, which is much higher than the typical velocity of DM in the Galactic halo. On the other hand, many theoretical models predict the existence of relativistic DM, such as cosmic ray boosted DM (CRDM)~\cite{Bringmann:2018cvk,Ema:2018bih,Cappiello:2019qsw, Dent:2019krz,Wang:2019jtk,Ge:2020yuf,Guo:2020oum,Xia:2020apm,Ema:2020ulo, Cao:2020bwd,Feng:2021hyz, Wang:2021nbf,Xia:2021vbz,PandaX-II:2021kai,CDEX:2022fig,Xia:2022tid,Maity:2022exk,Bell:2023sdq}, multi-component DM~\cite{Agashe:2014yua, Berger:2014sqa, Agashe:2015xkj, Su:2023zgr}, atmospheric DM~\cite{Alvey:2019zaa,Su:2020zny,Arguelles:2022fqq,Darme:2022bew,Su:2022wpj}, the solar reflection DM~\cite{An:2017ojc,Emken:2021lgc,An:2021qdl,Cappiello:2022exa}, and other possibilities~\cite{Oscura:2023qch}. These boosted DMs could produce plasmon signals in semi-conductor targets. Especially in the scenario with a light mediator, the plasmon resonance pole appears in the lower momentum transfer $Q$ region, which can further enhance the sensitivity. 
By using the first data from the silicon detector SENSEI at SNOLAB, we obtain the new bound on the DM-electron scattering cross section, which is about 3 to 20 times stronger than that from the XENON1T for DM masses ranging from 1~keV to 1~MeV.

{\it Collective excitation rate of boosted DM.} Among various relativistic DM models, we take the cosmic ray boosted DM as a benchmark, wherein a fraction of light DM in the halo is up-scattered by the energetic electrons in the cosmic rays. We focus on the interaction between DM ($\chi$) and electron ($e$) mediated by a dark vector ($A^{\prime}$). The relevant Lagrangian is given by $\mathcal{L}_{\mathrm{int}}  \supset g_{\chi}\bar{\chi}\gamma^{\mu}\chi A_{\mu}^{\prime}+g_{e}\bar{e}\gamma^{\mu}eA_{\mu}^{\prime}$, where $g_{\chi}$ and $g_{e}$ are the coupling constants for the dark vector-DM and dark vector-electron interactions, respectively. The flux of our CRDM on the Earth can be calculated by,
\begin{equation}
\begin{aligned}
\frac{\mathrm{d} \Phi_{\chi}}{\mathrm{d}  T_{\chi}} & = D_{\text {eff }} \frac{\rho_{\chi}^{\text {local }}}{m_{\chi}} \int_{T_{e}^{\min }}^{\infty} \mathrm{d}  T_{e} \frac{\mathrm{d} \sigma_{\chi e}}{\mathrm{d} T_{\chi}} \frac{\mathrm{d} \Phi_{e}^{\mathrm{LIS}}}{\mathrm{d}  T_{e}}.
\end{aligned}\label{eq:CReflux}
\end{equation}
Here $T_\chi$ and $T_e$ stand for the kinetic energies of the DM and the CR electron, respectively. In the DM rest frame, the minimal kinetic energy of electron is $T_{e}^{\min }=\left(\frac{T_{\chi}}{2}-m_{e}\right)\left[1 \pm \sqrt{1+\frac{2 T_{\chi}}{m_{\chi}} \frac{\left(m_{e}+m_{\chi}\right)^{2}}{\left(2 m_{e}-T_{\chi}\right)^{2}}}\right]$, where the signs $\pm$ correspond to $T_{\chi} > 2 m_e $ and $T_{\chi} < 2 m_e$, respectively. When $T_{\chi} = 2 m_e$,  $T_{e}^{\min }=(m_e+m_\chi)\sqrt{{m_e}/{m_\chi}}$. For 
details of CRDM kinematics, see Supplemental Material~\citep{supp_ref}.  $\rho_{\chi}^{\text{local}}=0.4$~GeV is the local DM halo density. The effective distance parameter is defined as $D_{\mathrm{eff}}\equiv \frac{1}{4 \pi \rho_{\chi}^{\text {local }}} \int_{\text {l.o.s }} \rho_{\chi} \mathrm{d} \ell \mathrm{d} \Omega$. We adopt the Navarro-Frenk-White DM profile $\rho_\chi$ with the scale radius $r_s =20$~kpc. Under the assumption of the homogeneous CR distribution in local interstellar (LIS), we obtain the effective distance $D_{\text {eff}} = 8.02~\text {kpc}$ after performing the full line-of-sight (l.o.s) integration over to 10~kpc~\cite{Bringmann:2018cvk}. A larger volume of the galactic CRs may be more realistic e.g., Ref.~\cite{Strong:1998fr}, where the inhomogeneous distribution of CRs and the dependence of $D_\mathrm{eff}$ on the DM kinetic energy needs to be considered~\cite{Xia:2021vbz, Xia:2022tid}. It leads to a larger value of $D_{\text {eff}}$, which would produce stronger limits because of the larger CRDM fux~\cite{Cappiello:2019qsw}. The cosmic electron spectrum in the LIS is taken as ${\mathrm{d} \Phi_{e}^{\mathrm{LIS}}}/{\mathrm{d}  T_{e}} = 4 \pi {\mathrm{d} I_e}/{\mathrm{d}  T_{e}}$, where the differential intensity ${\mathrm{d} I_e}/{\mathrm{d}  T_{e}}$ is detailed in Ref.~\cite{Boschini:2018zdv}. The uncertainties of the LIS population of CRs arise from the various propagation models and the injection spectrum parameters, such as the break rigidities~\cite{Strong:1998pw, Boschini:2018zdv}. For instance, the CR electron spectra can differ at most by about a factor of two for the different propagation models~\cite{Zhang:2023ajh}, which may change the resulting exclusion limits by a factor of $\sqrt{2}$, because the collective excitation event rate scales with cross section squared. In Eq.~(\ref{eq:CReflux}), the DM-CR electron scattering cross section is given by  
\begin{equation}
\begin{aligned}
\frac{d \sigma_{\chi e}}{d T_{\chi}}&=\frac{\bar{\sigma}_{\chi e} |F_{\mathrm{DM}}(q)|^2}{4 \mu_{\chi e}^{2} \left(2 m_{e} T_{e}+T_{e}^{2}\right)} \left[m_{\chi} T_{\chi}^{2} + 2 m_{\chi}\left(m_{e}+T_{e}\right)^{2}\right. \\
&\left.-T_{\chi}\left(m_{e}+m_{\chi}\right)^{2}-2 m_{\chi} T_{\chi}T_{e}\right],
\end{aligned}
\label{eq:diff_cross_section}
\end{equation}
with the DM form factor defined as 
\begin{equation}
\left|F_{\mathrm{DM}}(q)\right| \equiv (q^2_0 + m_{A^\prime}^2)/(q^2+m_{A^{\prime}}^2), 
\label{eq:formfactor}
\end{equation} 
where the four-momentum transfer squared $q^2 = 2 m_{\chi} T_{\chi}$ and the reference momentum $q_0 \equiv \alpha m_e \simeq3.7~\mathrm{keV}$. $\alpha$ is the electromagnetic fine structure constant. $m_{A^\prime}$ is the mediator mass and $\mu_{\chi e}$ is the reduced mass of the DM and the electron. By convention, we will show the sensitivity in terms of a momentum-independent reference cross section $\bar{\sigma}_{\chi e}  \equiv  \frac{\mu_{\chi e}^{2}}{\pi}\left(\frac{g_{\chi}g_{e}}{q^{2}_0+m_{A^\prime}^{2}}\right)^{2}$. Furthermore, we present our results for two scenarios: the light mediator case with $m^{2}_{A^\prime} \ll q^2$ where the DM form factor is $|F_{\mathrm{DM}}|^2 = (\alpha m_e)^4/q^4$, and the heavy mediator case with $m^{2}_{A^\prime} \gg q^2$, where the form factor simplifies to $|F_{\mathrm{DM}}|^2 =1$.

\begin{figure}[ht]
  \centering
\includegraphics[height=6cm,width=8cm]{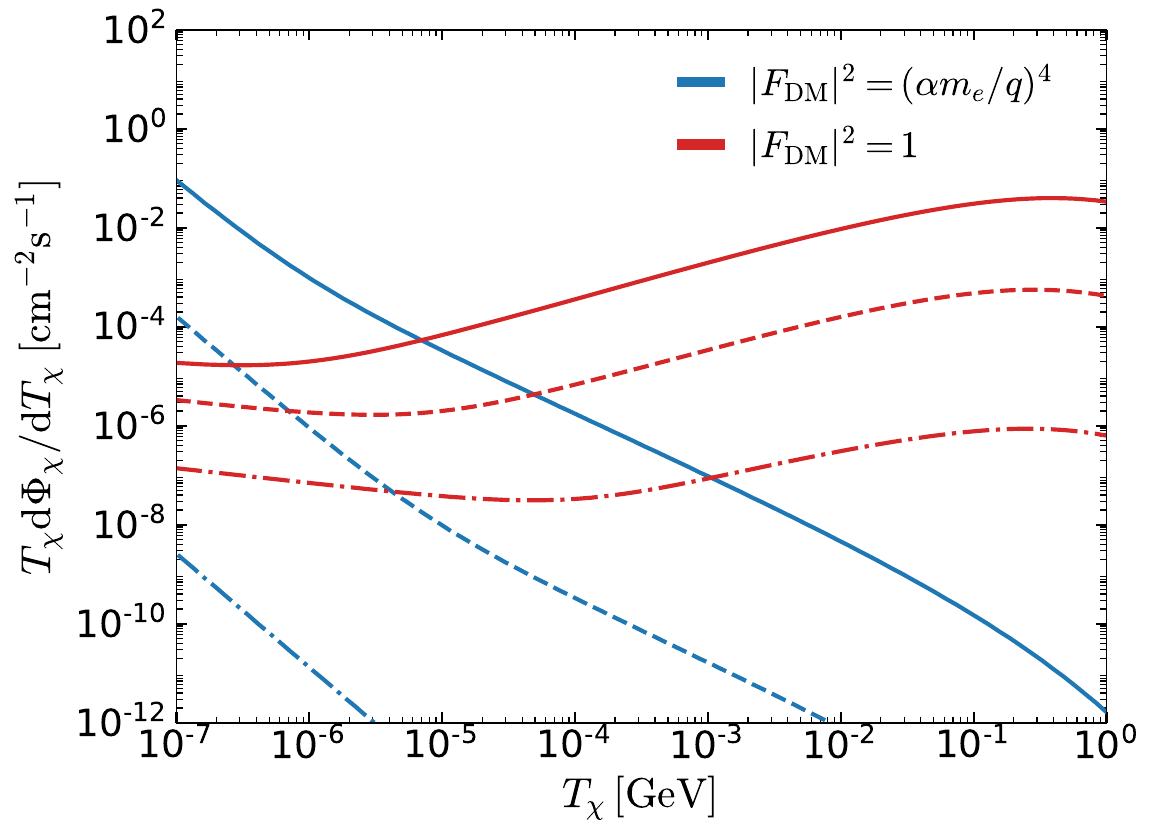}
\caption{The flux of CRDM as the function of DM kinetic energy $T_{\chi}$ from the DM-CR electron scattering via the light mediator $|F_{\mathrm{DM}}|^2 = (\alpha m_e)^4/q^4$ (blue lines) and the heavy mediator $|F_{\mathrm{DM}}|^2 =1$ (red lines). The reference cross section $\bar{\sigma}_{\chi e }  = 10^{-34} \; \mathrm{cm}^2$ and DM mass $m_{\chi} =1 $ keV (solid lines), $10$~keV (dashed lines), $m_e$~(dash-dotted lines) are adopted as the benchmark points.}
\label{fig:CRDM_flux}
\end{figure} 

\begin{figure*}[ht]
  \centering
\includegraphics[height=6cm,width=8cm]{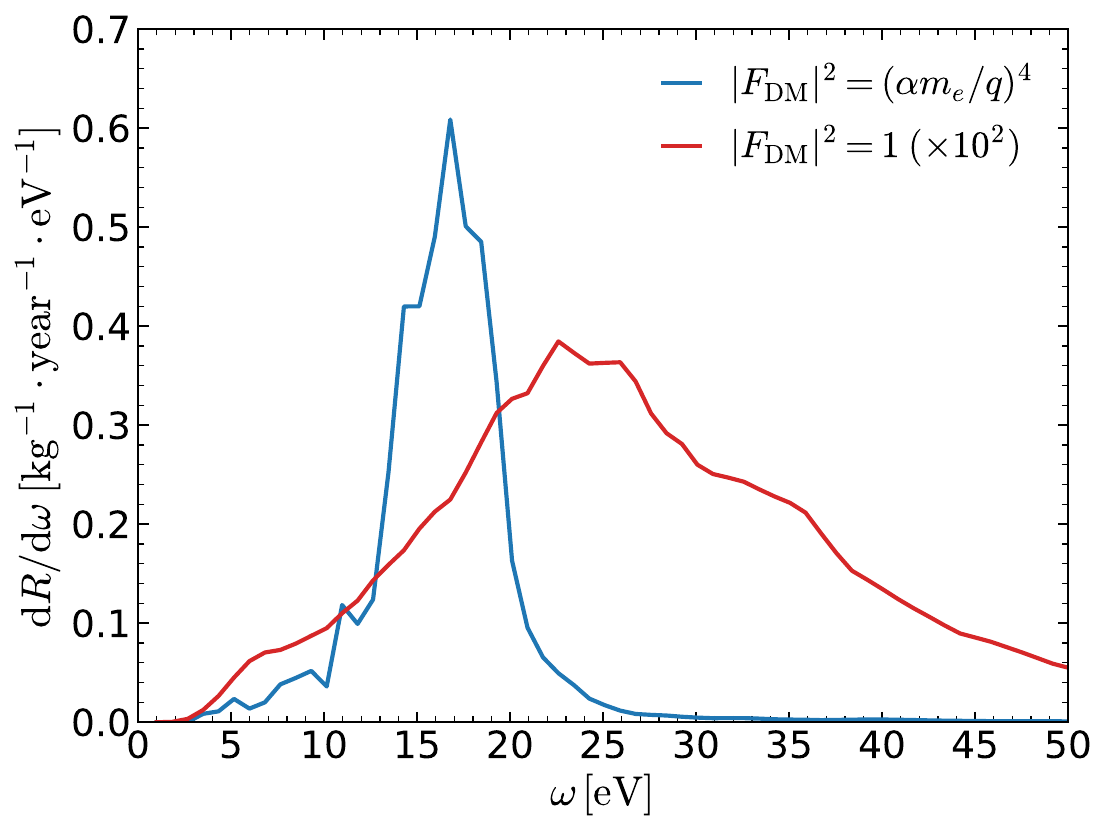}
\includegraphics[height=6cm,width=8cm]{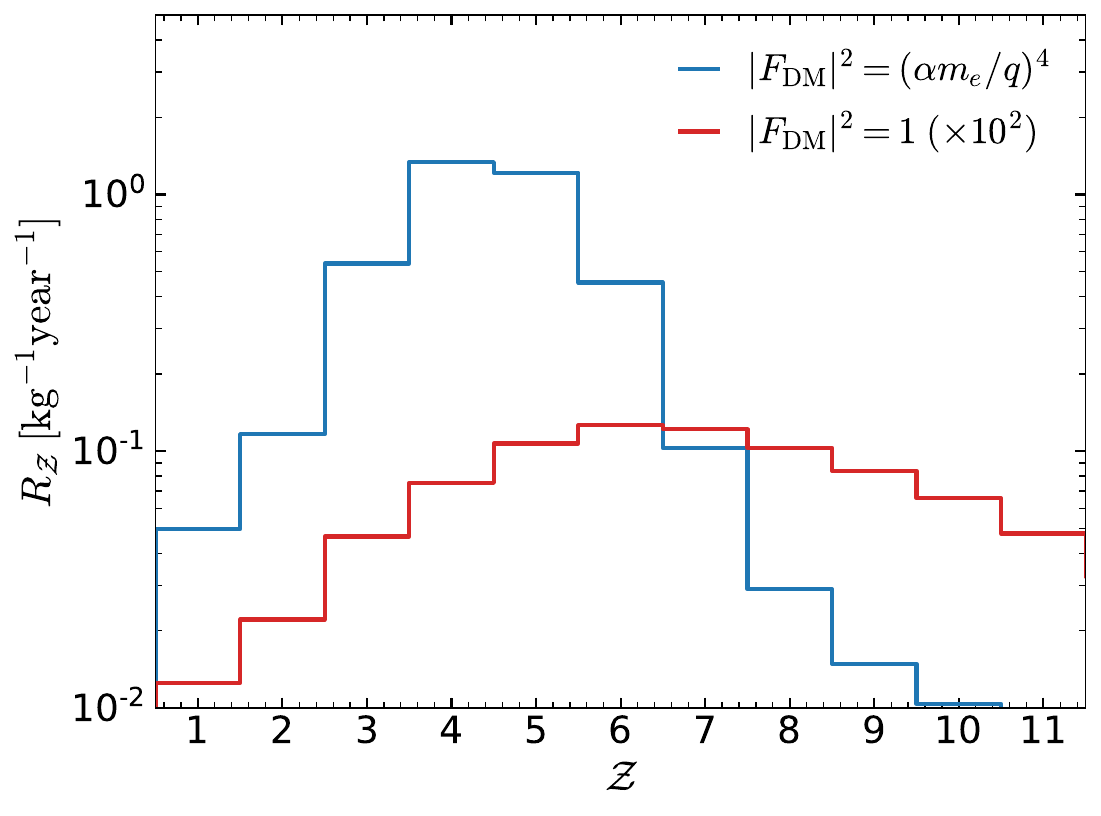}
\caption{ {\it Left panel:} differential event rate $\mathrm{d}R/\mathrm{d}\omega$ as a function of the deposited energy $\omega$; {\it Right panel:} event rate $R_{\mathcal{Z}}$ as a function of the ionized charge $\mathcal{Z}$. Here we assume the DM mass $m_\chi=1$ keV and the reference DM-electron scattering cross section $\bar{\sigma}_{\chi e} = 10^{-34} \; \mathrm{cm}^2$. The results are shown for both light (blue lines) and heavy (red lines) mediators. }
\label{fig:dRdomega}
\end{figure*} 
In Fig.~\ref{fig:CRDM_flux}, we present the CRDM flux produced by the scattering of the DM with the electron in the cosmic rays at the DM mass $m_\chi=$1~keV (solid lines), 10~keV (dashed lines), and $m_e$ (dotted lines), for both light (blue lines) and heavy (red lines) mediators. Compared to heavy mediators, the CRDM flux has a stronger dependence on the DM mass $m_\chi$ in the case of light mediators. This is because the $|F_{\mathrm{DM}}|^2$ term is proportional to $1/q^4 \propto 1/m_{\chi}^2 T_{\chi}^2$ for light mediators and remains constant for heavy mediators. As expected, the CRDM is boosted and obtains (semi-)relativistic velocities. We also compared our flux with that of Ref.~\cite{Cappiello:2019qsw} and found they are consistent by using the same parameters. If the interaction of the DM with the quarks is introduced, the DM can be accelerated by the proton in the cosmic rays as well. However, for a sub-MeV CRDM, the flux induced by the protons is much less than that induced by the electrons. This is because electrons are more efficient at transferring a larger portion of their energy to sub-MeV dark matter than protons~\cite{Cappiello:2019qsw}.

As aforementioned, such a (semi-)relativistic CRDM could obtain enough kinetic energy to excite the plasmons via the DM-electron scattering in semi-conductor detectors. In previous works, the energy loss function has been used to treat the non-relativistic DM-electron scattering. This approach incorporates many-body in-medium effects, which can be numerically implemented by using first-principles methods based on the DFT. We generalize the calculation framework~\cite{Hochberg:2021pkt} to obtain the electronic transition rate induced by a relativistic DM in the semi-conductor. For a DM with the mass $m_{\chi}$ and the momentum $\mathbf{p}_{\chi}$, the transition rate per unit volume is given by,
\begin{equation}
\Gamma\left(\mathbf{p}_{\chi}\right)=\int\frac{\mathrm{d}^{3}\mathbf{Q}}{\left(2\pi\right)^{3}}\left|V\left(\mathbf{Q},\omega\right)\right|^{2}\left[2\frac{Q^{2}}{e^{2}}\mathrm{Im}\left(-\frac{1}{\epsilon\left(\mathbf{Q},\omega \right)}\right)\right],   
\label{eq:Gamma}\end{equation}
where $e$ is the electron charge, $Q=\left|\mathbf{\mathbf{p}}_{\chi}-\mathbf{\mathbf{p}}_{\chi}'\right|$ and  $\omega \equiv E_{\chi}-E_{\chi}^{\prime}=\sqrt{p_{\chi}^{2}+m_{\chi}^{2}}-\sqrt{\left|\mathbf{p}_{\chi}-\mathbf{Q}\right|^{2}+m_{\chi}^{2}}$. For the isotropic target materials, the energy loss function can be simplified to $\mathrm{Im}\left[{-\epsilon^{-1}\left(\mathbf{Q},\omega\right)}\right]=\mathrm{Im}\left[{-\epsilon^{-1}\left(Q,\omega\right)}\right]$, and thus the transition rate becomes $\Gamma(\mathbf{p}_{\chi}) = \Gamma(p_{\chi})$. We utilize the energy loss function calculated by the \textit{DarkELF} package~\cite{Knapen:2021bwg}. For further details about the derivation of the transition rate, see Supplemental Material~\cite{supp_ref}. In Eq.~(\ref{eq:Gamma}), the potential $V\left(\mathbf{Q},\omega\right)$ for the interaction of the DM with the electron via a vector mediator is given by,
\begin{equation}
\begin{aligned}\left|V\left(\mathbf{Q},\omega\right)\right|^2 & =\frac{\pi\bar{\sigma}_{\chi e}[(2E_\chi-\omega)^2-Q^2]}{4\mu_{\chi e}^2E_\chi\left(E_\chi-\omega\right)} |F_{\mathrm{DM}}(q)|^2,
\label{eq:RelV}\end{aligned}   
\end{equation}
where the DM form factor is the same as that in Eq.~(\ref{eq:formfactor}) but with the four-momentum transfer squared $q^2 = -(\omega^2 - Q^2)$. In contrast with the non-relativistic case, the relativistic DM-electron potential $V\left(\mathbf{Q},\omega \right) $ is dependent on both the three-momentum transfer $\mathbf{Q}$ and the energy transfer $\omega$. It is straightforward to verify that this relativistic DM-electron interaction potential can be reduced to the non-relativistic one in the limit of $p_\chi\ll m_\chi$, i.e., $\left|V\left(\mathbf{Q},\omega \right)\right|^{2}\simeq g_{\chi}^{2}g_{e}^{2}/\left(Q^{2}+m_{A^{\prime}}^{2}\right)^{2}$.

By convoluting the CRDM flux in Eq.~(\ref{eq:CReflux}) with the transition rate in Eq.~(\ref{eq:Gamma}), we can have the differential event rate per unit volume,
\begin{equation}
\frac{\mathrm{d} R}{\mathrm{d} \omega} =\int \frac{\mathrm{d}T_{\chi}}{\rho_T} \int\frac{\mathrm{d} \Omega}{4 \pi}\frac{\mathrm{d}\Phi_{\chi}}{\mathrm{d}T_{\chi}} \left(\frac{E_{\chi}}{p_{\chi}}\right)\Gamma\left(p_{\chi}\right) \delta\left(E_{\chi}^{\prime}-E_{\chi}+\omega\right),
\label{eq:diffratio}
\end{equation}
where $\rho_T$ is the mass density of the semi-conductor target. The detailed calculations of the event rate are given in the Supplemental Material~\cite{supp_ref}.

\begin{figure*}[ht]
  \centering
\includegraphics[height=6cm,width=8cm]{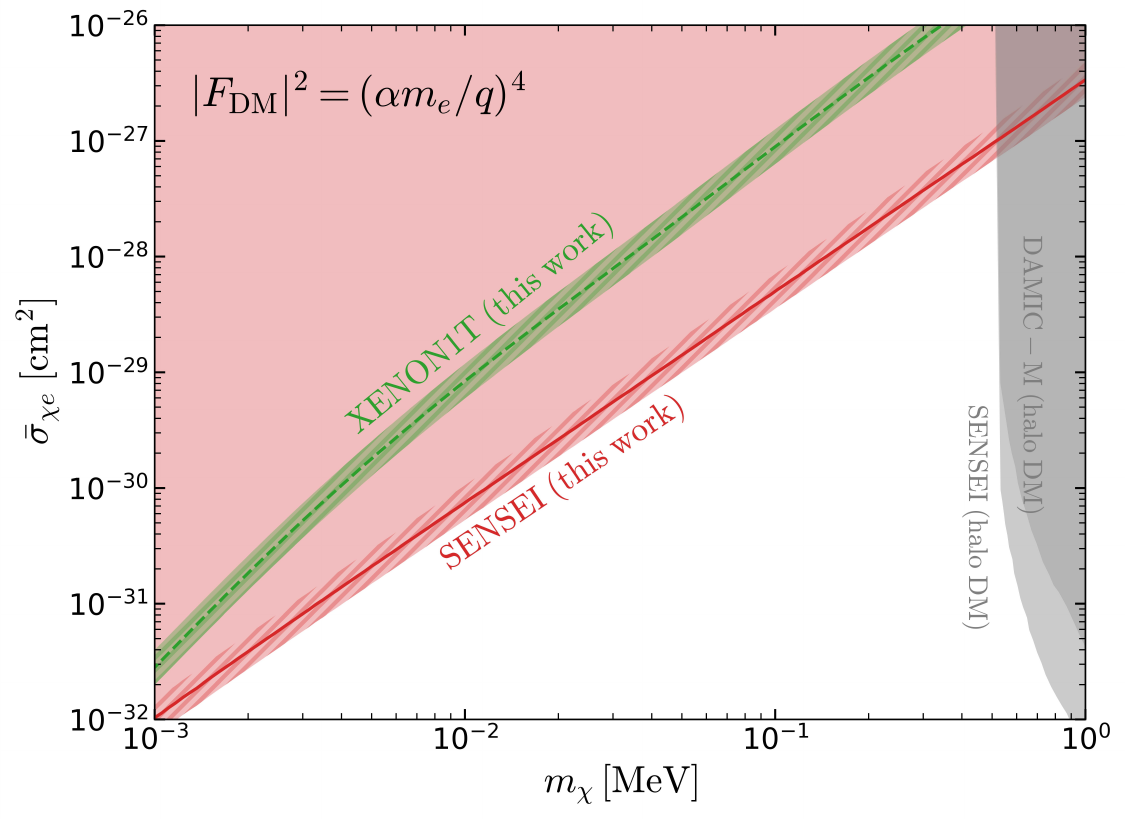}
\includegraphics[height=6cm,width=8cm]{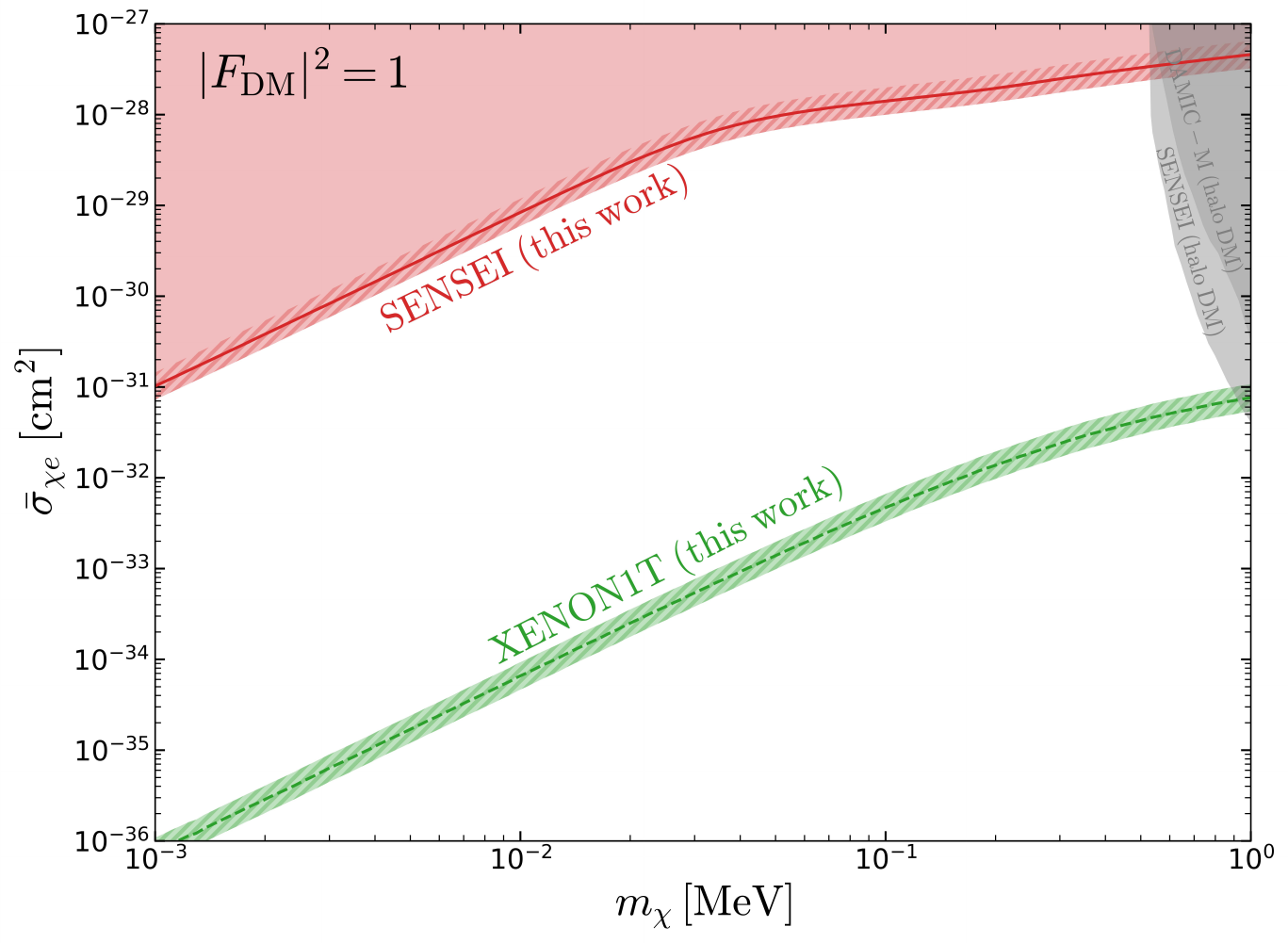}
\caption{The 90\% confidence level upper limits on the CRDM-electron scattering cross section $\bar{\sigma}_{\chi e }$ as a function of the DM mass for the light ({\it left panel}) and heavy ({\it right panel}) mediators. The regions above the red solid and green dashed lines are excluded by the null results of searches for DM from the SENSEI~\cite{SENSEI:2023zdf} and the XENON1T~\cite{xenon:collaboration} experiments, respectively. The hatched bands denotes the uncertainties on the limits from the LIS population of CRs~\cite{Zhang:2023ajh}. As the comparison, the constraints for the halo DM from DAMIC-M~\cite{DAMIC-M:2023gxo} and SENSEI~\cite{SENSEI:2020dpa} are plotted (gray shaded regions) as well. Besides, other constraints on halo
DM from PandaX-II~\cite{PandaX-II:2021nsg} and CDEX-10~\cite{CDEX:2022kcd} are for the DM mass
above $\mathcal{O}$(10) MeV, which is beyond the range we considered.} 
\label{fig:limit}
\end{figure*} 
{\it Exclusion limits.} We will present the sensitivity based on the silicon detector SENSEI with the skipper-CCDs at SNOLAB, which reported their results of searching for DM in a total exposure of 534.9 gram-days~\cite{SENSEI:2023zdf}. In the left panel of Fig.~\ref{fig:dRdomega}, we show the differential event rates for a 1~keV CRDM with the reference DM-electron scattering cross section $\bar{\sigma}_{\chi e} = 10^{-34} \; \mathrm{cm}^2$. As expected, the plasmon resonance appears in the low energy region for both light and heavy mediator scenarios. Such an effect is further enhanced by the DM form factor with light mediator. This results in the peak of the differential event rate with the light mediator being approximately two orders of magnitude larger than that with the heavy mediator. In the SENSEI experiment, the observable is the secondary electron-hole pairs.  Thus, we calculate the distribution of the event rate $R_{\mathcal{Z}}$ as a function of the ionized charge $\mathcal{Z}$ by using the secondary electron pair-creation probability distributions from the Ref.~\cite{Ramanathan:2020fwm}, where the band gap energy and mean energy-per pair are given by $E_{\mathrm{gap}} = 1.16\; \mathrm{eV}$ and $\epsilon= 3.75$ eV, respectively. The results are shown in the right panel of Fig.~3. We can see that the events for the light mediator mainly locate in the region of $ 3 \leq \mathcal{Z} \leq 6$, whereas events for the heavy mediator distribute in the higher $\mathcal{Z}$ region. Given the signal event distribution in Fig.~\ref{fig:dRdomega} and the low background reported in Ref.~\cite{SENSEI:2023zdf}, we use the data containing $3$ to $6$ electron-hole pairs with the corresponding event selection efficiencies~\cite{SENSEI:2023zdf} in our analysis. We summarize the experimental data from SENSEI and XENON1T that were used in our analysis, as well as the statistical methods for deriving the exclusion limits in the Supplementary Material~\cite{supp_ref}.

In Fig.~\ref{fig:limit}, we illustrate the exclusion limits on CRDM-electron scattering cross section $\bar{\sigma}_{\chi e }$ for the light and heavy mediators. In the light mediator scenario, the constraint from SENSEI on $\bar{\sigma}_{\chi e }$ in the DM mass range of 1 keV $<m_\chi<$ 1 MeV is approximately  3 to 20 times more stringent than that obtained from XENON1T with S2-only signal~\cite{xenon:collaboration}, because SENSEI can be accessible to lower momentum transfer that leads to an enhancement from the DM form factor. While in the heavy mediator scenario, the bound from SENSEI is much weaker than that from the XENON1T. This discrepancy arises due to the absence of DM form factor enhancement, and the larger detector volume of XENON1T. For a more detailed deviation of the XENON1T limit, see the Supplemental Material~\cite{supp_ref}. Furthermore, due to the lower threshold, the low energy DM can also contribute to the event rate in SENSEI, which is proportional to $1/v_{\chi}^2$. Therefore, when DM becomes heavy, the event rate from such low energy DM is enhanced. This effect is particularly evident in the heavy mediator scenario. Although such an enhancement also exists in the light mediator scenario, the slope of the exclusion limit is predominantly governed by the DM form factor.


{\it Conclusions.} We demonstrated that the nontrivial collective behavior of electrons in solid-state detectors, specifically plasmon excitation, is crucial for probing sub-MeV dark matter. We identified the conditions necessary for producing the plasmon resonance effect, requiring relativistic velocities for sub-GeV dark matter. We generalized the existing framework to deal with the plasmon excitation rate for relativistic dark matter, such as cosmic ray-boosted dark matter. Thanks to the recent silicon detector SENSEI with the skipper-CCDs at SNOLab, we use the data containing $3$ to $6$ electron-hole pairs to derive the exclusion limits on the CRDM-electron scattering cross section. We found that our limits from SENSEI on $\bar{\sigma}_{\chi e}$ in the dark matter mass range of 1 keV $< m_\chi <$ 1 MeV in the light mediator scenario can be about 3 to 20 times stronger than those from the XENON1T experiment. Our observation can apply to the investigations of other relativistic dark matter models, particularly within the unexplored sub-MeV mass range. Employing the plasmon resonance effect in forthcoming solid-state detectors, notably the upgraded SENSEI, DAMIC-M, and Oscura experiments, will offer a promising avenue for the detection of light dark matter.


\section{acknowledgments}
We thank Christopher V. Cappiello, Yan-Hao Xu, and Qiang Yuan for their helpful suggestions and discussions. This work is supported by the National Natural Science Foundation of China (NNSFC) under grants No. 12275134, No. 12275232, No. 12335005, and No. 12147228.

\bibliography{refs}







\onecolumngrid
\clearpage

\setcounter{page}{1}
\setcounter{equation}{0}
\setcounter{figure}{0}
\setcounter{table}{0}
\setcounter{section}{0}
\setcounter{subsection}{0}
\renewcommand{\theequation}{S.\arabic{equation}}
\renewcommand{\thefigure}{S\arabic{figure}}
\renewcommand{\thetable}{S\arabic{table}}
\renewcommand{\thesection}{\Roman{section}}
\renewcommand{\thesubsection}{\Alph{subsection}}
\newcommand{\ssection}[1]{
    \addtocounter{section}{1}
    \section{\thesection.~~~#1}
    \addtocounter{section}{-1}
    \refstepcounter{section}
}
\newcommand{\ssubsection}[1]{
    \addtocounter{subsection}{1}
    \subsection{\thesubsection.~~~#1}
    \addtocounter{subsection}{-1}
    \refstepcounter{subsection}
}
\newcommand{\fakeaffil}[2]{$^{#1}$\textit{#2}\\}

\thispagestyle{empty}
\begin{center}
    \begin{spacing}{1.2}
        \textbf{\large 
            \hypertarget{sm}{Supplemental Material:}
            \\Plasmon-enhanced Direct Detection Method for Boosted sub-MeV Dark Matter
        }
    \end{spacing}
    \par\smallskip
    Zheng-Liang Liang,$^{1}$
    Liangliang Su,$^{2}$
    Lei Wu,$^{2}$
    and Bin Zhu,$^{3}$
    \par
    {\small
        \fakeaffil{1}{College of Mathematics and Physics, Beijing University of Chemical Technology, Beijing, 100029, China}
        \fakeaffil{2}{Department of Physics and Institute of Theoretical Physics, Nanjing Normal University, Nanjing, 210023, China} 
        \fakeaffil{3}{Department of Physics, Yantai University, Yantai 264005, China}
        
    }

\end{center}
\par\smallskip

This Supplemental Material details the computational methods of a boosted DM scattering with the target material. It begins with a section on the kinematics of cosmic ray-boosted DM (CRDM), deriving the energy and momentum transfer in DM-electron scattering, including the minimal cosmic ray energy required for a given DM kinetic energy. Then, we present the collective excitation event rate of CRDM in the semi-conductor experiment, and the ionization event rate of CRDM in XENON1T experiment. Finally, we summarize the experimental data from SENSEI and XENON1T that were used in our analysis, as well as the statistical methods for deriving the exclusion limits.

\section{Kinematics of CRDM}

In the cosmic electron-DM scattering process, the initial DM particles are treated as being at rest, as their typical velocities ($v \sim 10^{-3}$) are negligible compared to the velocities of incoming cosmic electrons. In the DM rest frame (Lab frame), the total four-momentum of DM and cosmic electron is expressed as
\begin{equation}
\begin{array}{l}
p_e^\mu + p_\chi^\mu = (E, \boldsymbol{p}) = \left(E_e + m_\chi, \boldsymbol{p}_e\right), \\
\left(p_e^\mu + p_\chi^\mu\right)^2 = \left(E_e + m_\chi\right)^2 - \boldsymbol{p}_e^2 = \left(E_e^2 - \boldsymbol{p}_e^2\right) + m_\chi^2 + 2 E_e m_\chi = m_e^2 + m_\chi^2 + 2 E_e m_\chi,
\end{array}
\end{equation}
where $p_e^\mu$ represents the incoming cosmic electron four-momentum, and $p_\chi^\mu$ corresponds to the initial DM four-momentum. We also denote the outgoing electron by $p_e^{\prime\mu}$ and outgoing dark matter by $p_{\chi}^{\prime\mu}$. Now, boost $p_e^\mu$ to the center-of-mass frame (CM frame) along the $z$-axis, $p_e^\mu = \left(E_e, \boldsymbol{p}_e\right) = \left(E_e, 0, 0, p_{e}^z\right) \rightarrow p_e^\mu = \left(E_e^{\rm CM}, 0, 0, p_e^{z\,\rm CM }\right)$. We have
\begin{equation}
\begin{array}{l}
p_e^{z\,\rm CM} = \gamma\left(p_e^{z} - \beta E_e\right), \\
E_e^{\rm CM} = \gamma\left(E_e - \beta p_e^z\right),
\end{array}
\end{equation}
where
\begin{equation}
\begin{aligned}
\beta &= \frac{\boldsymbol{p}}{E} = \frac{\boldsymbol{p}_e}{E_e + m_\chi}, \\
\gamma &= \frac{E}{p} = \frac{E_e + m_\chi}{\sqrt{m_e^2 + m_\chi^2 + 2 E_e m_\chi}}.
\end{aligned}
\end{equation}
The momentum of the outgoing particle $p_e^{\prime\rm CM}$ is obtained by rotating $p_e^{\rm CM}$ by an angle $\theta_{\rm CM}$ in the center-of-mass frame:
\begin{equation}
\begin{array}{l}
p_e^{\prime 0\,\rm CM} = p_e^{0\,\rm CM}=\gamma\left(E_e - \beta p_e^z\right), \\
p_e^{\prime 1\,\rm CM} = p_e^{z\,\rm CM} \sin \theta_{\rm CM}, \\
p_e^{\prime 2\, \rm CM} = 0, \\
p_e^{\prime 3\, \rm CM} = p_e^{z\,\rm CM} \cos \theta_{\rm CM},
\end{array}
\end{equation}
where $\theta_{\rm CM}$ denotes the scattering angle in the center-of-mass frame. Boost this back to the Lab frame to obtain $p_e^{\prime}$, which involves applying the same Lorentz transformation but with a relative velocity $-\beta$:
\begin{equation}
E_e^{\prime} = E_e - \gamma^2 \beta\left(p_e^z - \beta E_e\right)(1 - \cos \theta_{\rm CM}).
\end{equation}

Thus, the resulting DM kinetic energy $T_\chi$ is determined by
\begin{equation}
T_\chi = T_e - T_e^{\prime} = E_e - E_e^{\prime} = \gamma^2 \beta\left(p_e^z - \beta E_e\right)(1 - \cos \theta_{\rm CM}) = \frac{m_\chi T_e\left(T_e + 2 m_e\right)}{\left(m_e + m_\chi\right)^2 + 2 T_e m_\chi}(1 - \cos \theta_{\rm CM}).
\label{eq:A1}
\end{equation}

For a given DM kinetic energy $T_{\chi}$, the minimal incoming CR energy can be determined by inverting Eq.~\ref{eq:A1}, 
\begin{equation}
    T_{e}^{\min }=\left(\frac{T_{\chi}}{2}-m_{e}\right)\left[1 \pm \sqrt{1+\frac{2 T_{\chi}}{m_{\chi}} \frac{\left(m_{e}+m_{\chi}\right)^{2}}{\left(2 m_{e}-T_{\chi}\right)^{2}}}\right], 
\end{equation}
where the signs $\pm$ correspond to $T_{\chi} > 2 m_e $ and $T_{\chi} < 2 m_e$, respectively. When $T_{\chi} = 2 m_e$,  $T_{e}^{\min }=(m_e+m_\chi)\sqrt{{m_e}/{m_\chi}}$.

\section{Event rate of CRDM scattering with the target materials}

{\it Collective Excitation Event Rate in the Semi-conductor Experiment.} We provide detailed calculations of the scattering rate in Eq.~(\ref{eq:Gamma}) and the CRDM-electron potential in Eq.~(\ref{eq:RelV}).  
For the DM-electron interaction, $\mathcal{L}_{\mathrm{int}}  \supset g_{\chi}\bar{\chi}\gamma^{\mu}\chi A_{\mu}^{\prime}+g_{e}\bar{e}\gamma^{\mu}eA_{\mu}^{\prime}$, the scattering rate $\Gamma(\mathbf{p}_{\chi})$ of an electron from the ground state $\ket{0}$ in semiconductor target induced by a relativistic DM particle with the momentum $\mathbf{p}_{\chi}$ can be obtained by the Fermi's golden rule or equivalently the Born approximation in the context of scattering theory,
\begin{equation}
\Gamma\left(\mathbf{p}_{\chi}\right)  = \sum_{f}\int\frac{\mathrm{d}^{3}p_{\chi}^{\prime}}{\left(2\pi\right)^{3}}\frac{\left|\mathcal{M}_{i\to f}\right|^{2}}{4E_{\chi}E_{\chi}^{\prime}}2\pi\delta\left(\omega_f+E_{\chi}^{\prime}-E_{\chi}\right),
\end{equation}
where $\omega_f$ is the energy of final state $\ket{f}$ of electron under the assumption of zero energy ground state. Meanwhile, the transfer energy is defined by the $\omega = E_{\chi}-E_{\chi}^{\prime}=\sqrt{p_{\chi}^{2}+m_{\chi}^{2}} -\sqrt{\left|\mathbf{p}_{\chi}-\mathbf{Q}\right|^{2}+m_{\chi}^{2}}$  with $p_{\chi}=\left|\mathbf{p_{\chi}}\right|$ and transfer momentum $\mathbf{Q}$. The amplitude $\mathcal{M}_{i\to f}$ is related to the $T$-matrix via
\begin{equation}
\begin{aligned}
\braket{\mathbf{p}_{\chi}^\prime,f|\,iT\,|\mathbf{p}_{\chi},0} & =  i\mathcal{M}_{i\rightarrow f}\,2\pi\delta\left(\omega_{f}+E_{\chi}^{\prime}-E_{\chi}\right) \\
 & =  \bar{u}_{p_{\chi}^{\prime}}^{s^{\prime}}\gamma^{0}u_{p_{\chi}}^{s}\,\frac{i\,g_{\chi}g_{e}}{q^{2}-\omega_{f}^{2}+m_{A^{\prime}}^{2}}\braket{f|\hat{\rho}(\mathbf{Q})|0}\,2\pi\delta\left(\omega_{f}- \omega\right),
\end{aligned}
\end{equation}
where $\hat{\rho}(\mathbf{Q})=\int\mathrm{d}^{3}\mathbf{x}\sum_{i}\delta(\mathbf{x}-\hat{\mathbf{r}}_{i})e^{-i\mathbf{Q}\cdot\mathbf{x}}=\sum_{i}e^{-i\mathbf{Q}\cdot\hat{\mathbf{r}}_{i}}$ is the momentum-space electron density operator, with $\left\{ \hat{\mathbf{r}}_{i}\right\} $ being the position operators of relevant electrons in the target. The energy loss function can be expressed in terms of the electron density operator as follows,
\begin{equation}
\mathrm{Im}\left(-\frac{1}{\epsilon\left(\mathbf{Q},\omega\right)}\right)=\frac{\pi e^{2}}{q^{2}}\sum_{f}\bigl|\langle f|\hat{\rho}(\mathbf{Q})|0\rangle\bigr|^{2}\delta\left(\omega_{f}-\omega\right),
\end{equation}
where $e^2 = 4 \pi \alpha$. Therefore, the scattering rate can be rewritten as

\begin{align}\label{eq:gamma}
\Gamma\left(\mathbf{p}_{\chi}\right) & = \sum_{f}\int\frac{\mathrm{d}^{3}p_{\chi}^{\prime}}{\left(2\pi\right)^{3}}\frac{\left|\mathcal{M}_{i\to f}\right|^{2}}{4E_{\chi}E_{\chi}^{\prime}}2\pi\delta\left(\omega_{f}+E_{\chi}^{\prime}-E_{\chi}\right) \notag \\
 & =  \int\frac{\mathrm{d}^{3}\mathbf{Q}}{\left(2\pi\right)^{3}}\left(\frac{1}{2}\sum_{s,s^{\prime}}\frac{\bar{u}_{p_{\chi}^{\prime}}^{s^{\prime}}\gamma^{0}u_{p_{\chi}}^{s}\bar{u}_{p_{\chi}}^{s}\gamma^{0}u_{p_{\chi}^{\prime}}^{s^{\prime}}}{4E_{\chi}E_{\chi}^{\prime}}\right)\,\left|\frac{i\,g_{\chi}g_{e}}{Q^{2}-\omega^{2}+m_{A^{\prime}}^{2}}\right|^{2}\sum_{f}\bigl|\langle f|\hat{\rho}(\mathbf{Q})|0\rangle\bigr|^{2}2\pi\delta\left(\omega_{f}-\omega \right)  \\
& =  \int\frac{\mathrm{d}^{3}\mathbf{Q}}{\left(2\pi\right)^{3}}\left(\frac{(2E_{\chi}-\omega)^{2}-Q^{2}}{4E_{\chi}(E_{\chi}-\omega)}\right)\left(\frac{g_{\chi}g_{e}}{Q^{2}-\omega^{2}+m_{A^{\prime}}^{2}}\right)^{2}\left[2\frac{Q^{2}}{e^{2}}\mathrm{Im}\left(-\frac{1}{\epsilon\left(\mathbf{Q},\omega\right)}\right)\right] \notag\\
&=   \int\frac{\mathrm{d}^{3}\mathbf{Q}}{\left(2\pi\right)^{3}} \left|V\left(\mathbf{Q}\right)\right|^{2} \left[2\frac{Q^{2}}{e^{2}}\mathrm{Im}\left(-\frac{1}{\epsilon\left(\mathbf{Q},\omega\right)}\right)\right], \notag
\end{align}

where the relativistic DM-electron potential in momentum space is defined as 
\begin{equation}
\begin{aligned}
\left|V\left(\mathbf{Q}\right)\right|^{2} &= \left(\frac{(2E_{\chi}-\omega)^{2}-Q^{2}}{4E_{\chi}(E_{\chi}-\omega)}\right)\left(\frac{g_{\chi}g_{e}}{Q^{2}-\omega^{2}+m_{A^{\prime}}^{2}}\right)^{2} \\ 
& = \frac{\pi\overline{\sigma}_{\chi e}[(2E_\chi-\omega)^2-Q^2]}{4\mu_{\chi e}^2E_\chi\left(E_\chi-\omega\right)} |F_{\mathrm{DM}}(q)|^2,
\end{aligned}
\end{equation}
with 
\begin{equation}
    \left|F_{\mathrm{DM}}(q)\right|^{2} = \frac{(q^2_0 + m_{A^\prime}^2)^2}{(q^2+m_{A^{\prime}}^2)^2}, \;\;\; \bar{\sigma}_{\chi e}  \equiv  \frac{\mu_{\chi e}^{2}}{\pi}\left(\frac{g_{\chi}g_{e}}{q^{2}_0+m_{A^\prime}^{2}}\right)^{2},
\end{equation}
where $q^2_0=\alpha^2 m^2_e$ and $q^2 = Q^2 -\omega^2$. It is straightforward to verify that the relativistic DM-electron potential is reduced to the nonrelativistic one in the limit $p_\chi\ll m_\chi$, i.e., $\left|V\left(\mathbf{Q},\omega \right)\right|^{2}\simeq g_{\chi}^{2}g_{e}^{2}/\left(Q^{2}+m_{A^{\prime}}^{2}\right)^{2}$ as shown in Ref.~\cite{Hochberg:2021pkt}.

With Eq.~\ref{eq:gamma}, we can obtain the collective excitation event rate in semiconductors induced by a relativistic DM flux. Because the semiconductor target can be regarded as isotropic (i.e., $\mathrm{Im}\left[-\epsilon^{-1}\left(Q,\omega\right)\right]$ is independent of the direction of $\mathbf{Q}$), we take the direction of initial momentum $\mathbf{p}_{\chi}$ as the $z$-axis of the spherical coordinate system and integrate out the polar angle of $\mathbf{Q}$ to $\mathbf{\mathbf{p}_{\chi}}$, so we have
\begin{align}
\label{eq:event_rate}
\frac{\mathrm{d} R}{\mathrm{d} \omega} & = \frac{1}{\rho_T}\int\mathrm{d}T_{\chi} \int\frac{\mathrm{d} \Omega}{4 \pi} \frac{\mathrm{d}\Phi_{\chi}}{\mathrm{d}T_{\chi}}\left(\frac{E_{\chi}}{p_{\chi}}\right)\Gamma\left(p_{\chi}\right) \delta\left(E_{\chi}^{\prime}-E_{\chi}+\omega\right) \notag \\
& = \frac{1}{\rho_T}\frac{2\pi\bar{\sigma}_{\chi e}}{\mu_{\chi e}^{2}} \int\mathrm{d} E_{\chi}\frac{\mathrm{d}\Phi_{\chi}}{\mathrm{d}T_{\chi}}\int\frac{Q^{2}\mathrm{d}Q\mathrm{d}\cos\theta_{\mathbf{Q}\mathbf{p}_{\chi}}\mathrm{d}\phi}{\left(2\pi\right)^{3}}\,\frac{(E_{\chi}+E_{\chi}^{\prime})^{2}-Q^{2}}{4E_{\chi}E_{\chi}^{\prime}}\left|F_{\mathrm{DM}}(q)\right|^{2} \notag \\
 & \times\frac{Q^{2}}{e^{2}}\frac{E_{\chi}}{p_{\chi}}\mathrm{Im}\left[\frac{-1}{\epsilon\left(Q,\omega\right)}\right]\,\left.\delta\left(E_{\chi}^{\prime}-E_{\chi}+\omega\right)\right|_{E_{\chi}^{\prime}=\sqrt{\left|\mathbf{p}_{\chi}-\mathbf{Q}\right|^{2}+m_{\chi}^{2}}} \notag \\
 & = \frac{1}{\rho_T}\frac{\bar{\sigma}_{\chi e}}{e^{2}\mu_{\chi e}^{2}} \int\frac{Q^{2}\mathrm{d}Q}{2\pi}\int\mathrm{d}E_{\chi}\frac{\mathrm{d}\Phi_{\chi}}{\mathrm{d}T_{\chi}}\int\left|\frac{\mathrm{d}\cos\theta_{\mathbf{Q}\mathbf{p}_{\chi}}}{\mathrm{d}E_{\chi}^{\prime}}\right|\mathrm{d}E_{\chi}^{\prime}\frac{Q^{2}E_{\chi}}{p_{\chi}}\frac{(E_{\chi}+E_{\chi}^{\prime})^{2}-Q^{2}}{4E_{\chi}E_{\chi}^{\prime}} \notag \\
&\times\left|F_{\mathrm{DM}}(q)\right|^{2}\mathrm{Im}\left[\frac{-1}{\epsilon\left(Q,\omega\right)}\right]\,\delta\left(E_{\chi}^{\prime}-E_{\chi}+\omega\right) \\
&=\frac{1}{\rho_T}\frac{\bar{\sigma}_{\chi e}}{2\pi e^{2}\mu_{\chi e}^{2}} \int Q^{2}\mathrm{d}Q\int\mathrm{d}E_{\chi}\frac{\mathrm{d}\Phi_{\chi}}{\mathrm{d}T_{\chi}}\int\mathrm{d}E_{\chi}^{\prime}\left(\frac{E_{\chi}E_{\chi}^{\prime}Q}{p_{\chi}^{2}}\right)\frac{(E_{\chi}+E_{\chi}^{\prime})^{2}-Q^{2}}{4E_{\chi}E_{\chi}^{\prime}} \notag\\
& \times\left|F_{\mathrm{DM}}(q)\right|^{2}\mathrm{Im}\left[\frac{-1}{\epsilon\left(Q,\omega\right)}\right]\,\delta\left(E_{\chi}^{\prime}-E_{\chi}+\omega\right) \notag \\
& = \frac{1}{\rho_T}\frac{\bar{\sigma}_{\chi e}}{2\pi e^2\mu_{\chi e}^{2}} \int Q^{3}\mathrm{d}Q\int\mathrm{d}E_{\chi}\frac{\mathrm{d}\Phi_{\chi}}{\mathrm{d}T_{\chi}}\frac{(2E_{\chi}-\omega)^{2}-Q^{2}}{4 v_{\chi}^2 E_{\chi}^{2}}  \left|F_{\mathrm{DM}}(q)\right|^{2} \notag\\
&\times \mathrm{Im}\left[\frac{-1}{\epsilon\left(Q,\omega\right)}\right]\,\Theta\left[E_{\chi}-E^{-}-\omega\right]  \notag,
\end{align}
where $\rho_T$ is the mass density of semiconductor target and $E^{\pm}\left(p_{\chi}\right)=\sqrt{\left(p_{\chi}\pm q\right)^{2}+m_{\chi}^{2}}$. In the first line, we insert an identity $1=\int\mathrm{d}\omega\,\delta\left(E_{\chi}^{\prime}-E_{\chi}+\omega\right)$ in order to introduce the variable of deposited energy $\omega$ and the $\mathrm{d} T_{\chi} = \mathrm{d} (E_{\chi}-m_{\chi}) =\mathrm{d} E_{\chi}$ is considered in second line. Then we take a variable transformation from $\cos\theta_{\mathbf{Q}\mathbf{p}_{\chi}}$ to $E_{p_{\chi}^{\prime}}$, along with its corresponding Jacobian 
\begin{eqnarray}
\left|\frac{\mathrm{d}\cos\theta_{\mathbf{Q}\mathbf{p}_{\chi}}}{\mathrm{d}E_{\chi}^{\prime}}\right|	=	\left(\frac{E_{\chi}^{\prime}}{p_{\chi}Q}\right),
\end{eqnarray} and the step function in the last line indicates whether the span of integration over $E_{p_{\chi}^{\prime}}$ covers the zero in the delta
function. In addition, if the DM flux is rewritten with the speed distribution function, $\int\mathrm{d}T_{\chi}\mathrm{d}\Phi_{\chi}/\mathrm{d}T_{\chi}=\left(\rho_{\chi}/m_{\chi}\right)\int\mathrm{d^{3}}v\,v_{\chi}f_{\chi}\left(\mathbf{v}_{\chi}\right)$, the Eq.~(\ref{eq:event_rate}) can spontaneously reduce to the form in Ref.~\cite{Knapen:2021run,Hochberg:2021pkt} for a nonrelativistic DM flux. Therefore, the event rate for the ionized charge $\mathcal{Z}$ can be given by 
$R_{\mathcal{Z}} = \int \mathrm{d} \omega P(\mathcal{Z}|\omega) {\mathrm{d} R}/{\mathrm{d} \omega}$, where $P(\mathcal{Z}|\omega)$ is the secondary electron pair-creation probability distributions in Ref.~{\cite{Ramanathan:2020fwm}}.

{\it Ionization Event Rate in XENON1T Experiment.} In XENON1T experiment, the differential electron ionization event rate induced by CRDM-electron scattering is given by 
\begin{equation}
\begin{aligned}
 \frac{\mathrm{d} R_{\mathrm{ion}}}{\mathrm{d} \ln \Delta E_e}&=N_{T} \frac{\bar{\sigma}_{e}}{8 \mu_{\chi e}^{2}} \int  \mathrm{d} Q \int_{T_{\min }} \mathrm{d} T_{\chi} \left|f_{\mathrm{i o n}}^{i}\left(k^{\prime}, Q\right)\right|^{2}   Q \frac{\mathrm{d} \Phi_{\chi}}{\mathrm{d} T_{\chi}}\\
 &\times \left|F_{\mathrm{DM}}(q)\right|^{2} \frac{8 m_{e}^2 E_{\chi}^{2}+q^4-2 q^2\left( m_{\chi}^2+m_e^2+2m_eE_{\chi}\right)}{8 m_{e}^2 v_{\chi}^2 E_{\chi}^2}
\end{aligned}
\label{eq:xenon_rate}
\end{equation}
Where $N_T$ is the number density of target nuclei. The DM form factor $|F_{\mathrm{DM}}(q)|^2$ is the same as that in Eq.~(\ref{eq:formfactor}) but with the four-momentum transfer squared $q^2 = -(\Delta E_e^2 - Q^2)$. The deposit energy $\Delta E_e = E_R+ |E_{nl}|$ is the function of the binding energy of $(n,l)$ shell $E_{nl}$ and the final electron energy $E_R = k^{\prime 2}/2 m_e$.  $\left|f_{\mathrm{i o n}}^{i}\left(k^{\prime}, Q\right)\right|^{2}$ is the ionization form factor, which describes an electron with initial state $(n,l)$ excited the continuum state with momentum $k^{\prime}$, and we adopted the results of Ref.~\cite{Flambaum:2020xxo,Catena:2019gfa} in this work. $T_{\min}$ is the minimum DM kinetic energy to obtain an electron with $E_R$ by DM-electron scattering, 
\begin{equation}
    p_{\chi}^{\min} = \frac{Q}{2\left(1-\Delta E_{e}^{2} / Q^{2}\right)}\left(1-\frac{\Delta E_{e}^{2}}{Q^{2}}+\frac{\Delta E_{e}}{Q} \sqrt{\left(1-\frac{\Delta E_{e}^{2}}{Q^{2}}\right)\left(1-\frac{\Delta E_{e}^{2}}{Q^{2}}+\frac{4 m_{\chi}^{2}}{Q^{2}}\right)}\right).
\end{equation}
The Eq.~(\ref{eq:xenon_rate}) can spontaneously reduce to the form in Ref.~\cite{Essig:2011nj,Essig:2015cda}, and more details can be found in the supplemental material in Ref.~\cite{Flambaum:2020xxo}. In general, the experimental observable is the number of photoelectrons (PE), and we need to make a transform, i.e.,
\begin{equation}
 R_{\mathrm{S2}}=\int \mathrm{d} \Delta E_{e} \epsilon(\mathrm{S} 2) P\left(\mathrm{~S} 2 \mid \Delta E_e\right) \frac{\mathrm{d} R_{\mathrm{i o n}}}{ \mathrm{d} \Delta E_{e}},
\end{equation}
where $\epsilon({\rm S2})$ is the detector efficiency. $P\left({\rm S2} \mid \Delta E_e\right)$ represents the probability function that converts deposited energy into the PE numbers for the S2 signal, as well as their product is tabulated by XENON1T collaboration in Ref.~\cite{xenon:collaboration}. 

{\section{Data and analysis method}}

In Table~\ref{tab:data}, we present the data used in our analysis, including the SENSEI and the XENON1T experiments.

\begin{table}[ht]
\begin{tabular}{c|cccc}
\hline
Experiment              & ROI bin                         & Event & Background & Exposures    \\ \hline
{SENSEI~\cite{SENSEI:2023zdf}} & $\mathcal{Z} = 3 e^{-}$                       & 4     & 0.07      & 57.71 $\mathrm{gram}\cdot \mathrm{days}$  \\ \cline{2-5} 
                        & $\mathcal{Z} = 4 e^{-}$                        & 0     & 0          & 63.03 $\mathrm{gram}\cdot \mathrm{days}$   \\ \cline{2-5} 
                        & $\mathcal{Z} = 5 e^{-}$                        & 0     & 0          & 65.56 $\mathrm{gram}\cdot \mathrm{days}$  \\ \cline{2-5} 
                        & $\mathcal{Z} = 6 e^{-}$                         & 0     & 0          & 67.73  $\mathrm{gram}\cdot \mathrm{days}$ \\ \hline
Xenon1T~\cite{xenon:collaboration}                & $\mathrm{PE}\in [165.3, 271.7]$ & 24.6  & $/$          & 0.97678  $\mathrm{tonne}\cdot \mathrm{year}$ \\ \hline
\end{tabular}
\caption{The data used in this work are from the SENSEI and the Xenon1T experiments. $\mathcal{Z}$ denotes the ionized charge. PE stands for the photoelectron counts.}
\label{tab:data}
\end{table}

In our analysis, we calculate the 90\% C.L. exclusion limit from SENSEI experiment by using the maximum log-likelihood function, assuming that the data follow a Poisson distribution. The combined likelihood function can be written as
\begin{equation}
    L\left(\bar{\sigma}_{\chi e} ; b\right)= \prod_{i} \frac{\left( s_{i}(\bar{\sigma}_{\chi e})+b_{i}+k_{i}\right)^{n_{i}}}{n_{i}!} \exp \left[-\left(s_{i}(\bar{\sigma}_{\chi e})+b_{i}+k_{i}\right)\right],
\end{equation}
where $n_i$ and $s_i$ are the observed and expected event numbers for bin $i$, respectively. The expected event number $s_i$ is a function of the dark matter-electron scattering cross section $\bar{\sigma}_{\chi e}$. $k_i$ represents the known background event number, and $b_i$ is a nuisance parameter denoting the unknown background event number. To test a hypothesized value of $\bar{\sigma}_{\chi e}$, the profile likelihood ratio is defined as
\begin{equation}
    \lambda(\bar{\sigma}_{\chi e})=\frac{L(\bar{\sigma}_{\chi e} ; \hat{\hat{\mathbf{b}}})}{L(\hat{\bar{\sigma}}_{\chi e}; \hat{\mathbf{b}})},
\end{equation}
where $\hat{\hat{\mathbf{b}}}$ is obtained by maximizing the likelihood function $L$ for the specified value of $\bar{\sigma}_{\chi e}$, and $\hat{\bar{\sigma}}_{\chi e}$ and $\hat{\mathbf{b}}$ are the corresponding maximum likelihood estimators. To obtain the 90\% C.L. upper limit, we consider the alternative test statistic $\tilde{q}_{\bar{\sigma}_{\chi e}}$, as described in Ref.~\cite{Cowan:2010js}:
\begin{equation}
    \tilde{q}_{\bar{\sigma}_{\chi e}}=\left\{\begin{array}{ll}
-2 \ln {\lambda}(\bar{\sigma}_{\chi e}) & \hat{\bar{\sigma}}_{\chi e} \leq \bar{\sigma}_{\chi e} \\
0 & \hat{\bar{\sigma}}_{\chi e}>\bar{\sigma}_{\chi e}
\end{array}\right..
\end{equation}
Next, we set the probability density function of $\tilde{q}_{\bar{\sigma}_{\chi e}}$ under the assumption of $\bar{\sigma}_{\chi e}$ equals 0.1, i.e., $f(\tilde{q}_{\bar{\sigma}_{\chi e}} | \bar{\sigma}_{\chi e}^{\text{limit}}) = 0.1$, to obtain the 90\% C.L. upper limit. In this work, we utilize the \textit{pyhf} package~\cite{pyhf,pyhf_joss} to perform these calculations. Since XENON1T collaboration has provided 90\% C.L. background-subtracted data, we can set the exclusion limits by requiring $N_{\rm th}/N_{\rm obs}>1$ as described in~\cite{xenon:collaboration}. Here $N_{\rm th}$ is the event number predicted by theory and $N_{\rm obs}$ is the observed event number.

\end{document}